# Observation of bound states in the continuum in a micromechanical resonator

Yue Yu, Xiang Xi, and Xiankai Sun[*]

*Department of Electronic Engineering, The Chinese University of Hong Kong, Shatin, New Territories, Hong Kong*

[*]*Corresponding author:* xksun@cuhk.edu.hk

**Bound states in the continuum (BICs) refer to physical states that possess intrinsic zero dissipation loss even though they are located in the continuous energy spectrum. BICs have been widely explored in optical and acoustic structures, leading to applications in sensing and lasing. Realizing BICs in micromechanical structures is of significant importance for both fundamental research and engineering applications. Here, we fabricated, with CMOS-compatible processes on a silicon chip, a wheel-shaped micromechanical resonator, in which we experimentally observed the BIC in the micromechanical domain. Such BICs result from destructive interference between two dissipative modes of the mechanical structure under broken azimuthal symmetry. These BICs are found to be robust against size variations of the dissipation channels. The demonstrated mechanical BIC can be obtained with a large and robust supporting structure, which substantially reduces device fabrication difficulty and allows for its operation in versatile environments for broader application areas. Our results open a new way of phonon trapping in micromechanical structures with dissipation channels, and produce long phonon lifetimes that are desired in many mechanical applications such as mechanical oscillators, sensors, and quantum information processors.**

**Introduction**

Micro- and nanomechanical resonators, which possess a very small mass and can be strongly coupled to light and matter, have been explored for precision metrology applications like mass and force sensing[1] and employed for investigating macroscopic quantum physics[2,3]. Reducing mechanical dissipation is crucial to these applications since it allows enhanced mechanical fields with long coherence time and thus leads to improved performance. The conventional wisdom of reducing the dissipation loss relies on separating their eigenmodes from the continuum of lossy modes by constructing deep energy potentials with different materials or periodic structures[4,5]. For another type of nonperiodic individual resonators, where the bandgap shielding strategy cannot be applied, reducing the dissipation loss relies on minimizing their supporting structure[6,7], which increases device fabrication difficulty and sets restrictions on their application areas. For example, devices based on such delicate mechanical structures cannot be used repeatedly for fluid-based applications, because they would likely fail when the ambient environment changes from a liquid to a gas.

Bound states in the continuum (BICs) refer to a type of eigenstates with infinite lifetime yet spectrally overlapping with lossy states in the continuum[8]. Originally introduced to quantum mechanics, the concept of BICs has been extended to optical[9-13] and mechanical[14,15] domains, and enabled many unprecedented applications such as low-threshold lasing[16-18], ultrasensitive sensing[19], and vortex beam generation[20]. To date, most experimental demonstrations of BICs in optics and mechanics are based on periodic structures[21,22] with certain symmetry. These devices usually have a large footprint with a large modal volume or effective mass, which sets limitations to their application scenarios. In contrast to devices based on periodic structures, nonperiodic individual optical and mechanical resonators can have more confined fields with stronger modal intensity at the micro/nanoscale, leading to a series of applications in precision metrology as well as studies of macroscopic quantum physics. BICs in an individual optical resonator have been demonstrated, producing efficient second-harmonic generation in nanoscale subwavelength dielectric cylinders[23]. However, experimental demonstration of BICs in an individual mechanical resonator remains elusive.

Here, we experimentally demonstrated mechanical BICs in a micromechanical resonator. By breaking the azimuthal symmetry, we introduced coupling between a radial-contour mode and a wine-glass mode of a wheel-shaped structure to obtain destructive interference between energy dissipation of the two modes, which produces a mechanical BIC under the Friedrich–Wintgen condition[24]. The mechanical BIC was experimentally confirmed by optomechanical measurement of the devices in a vacuum. In contrast to conventional BICs requiring certain symmetry, the demonstrated mechanical BICs represent a new paradigm for constructing high-$Q$ micromechanical resonators through symmetry breaking. As we could find the low-loss mechanical BIC for a wide range of the supporting rods' width from hundreds of nanometers to several micrometers, the high tolerance on the supporting rods' width substantially alleviates device fabrication difficulty and enables repeated operation in fluid-based applications. These results open a new way of phonon trapping in micromechanical structures with dissipation channels, and will bring impacts to the fields of electromechanics, optomechanics, and quantum physics.

## Results

### Friedrich–Wintgen BICs in a micromechanical resonator

To construct BICs in an individual mechanical resonator, suppose we have a resonator supporting two dissipative modes coupled to each other, as shown in Fig. 1a. Such a system can be described by a Hamiltonian

$$H = \begin{pmatrix} \omega_1 - i\gamma_1 & \kappa - i\sqrt{\gamma_1\gamma_2} \\ \kappa - i\sqrt{\gamma_1\gamma_2} & \omega_2 - i\gamma_2 \end{pmatrix}, \quad (1)$$

where $\omega_1$ ($\gamma_1$) and $\omega_2$ ($\gamma_2$) are the resonant frequencies (dissipation rates) of the two modes. The two modes are coupled with each other with a coupling coefficient $\kappa$, which results in an anticrossing of these two modes. At this anticrossing point, when the Friedrich–Wintgen condition

$$\kappa(\gamma_1 - \gamma_2) = \sqrt{\gamma_1\gamma_2}(\omega_1 - \omega_2) \quad (2)$$

is satisfied[24], the complex resonant frequencies become (See Supplementary Note 1):

$$\Omega_1 = \frac{\omega_1 + \omega_2}{2} + \frac{\kappa(\gamma_1 + \gamma_2)}{2\sqrt{\gamma_1\gamma_2}} - i(\gamma_1 + \gamma_2), \quad (3)$$

$$\Omega_2 = \frac{\omega_1 + \omega_2}{2} - \frac{\kappa(\gamma_1 + \gamma_2)}{2\sqrt{\gamma_1\gamma_2}}. \quad (4)$$

Equation (4) shows that the lower-frequency mode is a BIC because its complex resonant frequency has a vanishing imaginary part, which means it experiences zero dissipation loss. It should be noted that in this system, when one of the two hybrid modes becomes lossless, the Friedrich–Wintgen condition is always satisfied (See Supplementary Note 2). Therefore, one can verify a Friedrich–Wintgen BIC by measuring the dissipation loss of the two hybrid modes of the system.

First, we consider a ring-shaped thin-plate micromechanical resonator as shown in Fig. 1b. It is made in 220-nm-thick silicon and has an inner radius $r$ and an outer radius $R$ ($r$, $R \gg$ 220 nm). Such resonators support two types of in-plane mechanical modes: radial-contour modes and wine-glass modes. Since these two types of modes have different dependence on $r$, fixing $R$ = 26.1 μm and varying $r$ lead to a crossing of resonant frequencies of the fundamental radial-contour mode (mode A in Fig. 1b) and the 4th-order wine-glass mode (mode B in Fig. 1b). In a ring-shaped resonator with perfect azimuthal symmetry, mode A and mode B are orthogonal to each other without modal coupling. Therefore, the Friedrich–Wintgen condition in Eq. (2) for BICs cannot be satisfied. To introduce modal coupling for satisfying the Friedrich–Wintgen condition, we break the azimuthal symmetry of the ring-shaped resonator by modifying its inner boundary to an ellipse, with semi-major and semi-minor axes being respectively $r_x$ and $r_y$, as shown in Fig. 1c. Figure 1c also plots the simulated modal frequencies of the modified structure as a function of $r_x$ with fixed $r_y$ = 18.7 μm and $R$ = 26.1 μm, where an anticrossing occurs near $r_x$ = 20.6 μm indicating the coupling between the two mechanical modes. At the anticrossing point, the energy exchange between the original mode A and mode B leads to two hybrid modes, namely mode A′ and mode B′, as shown in Fig. 1d. Note that although the structure in Fig. 1c can have the required coupling between different modes which can support a BIC, a realistic device must also include supporting structures that are connected to the substrate, which act as the dissipation channel for both mechanical modes. Compared with the original modes A and B, the hybrid modes A′ and B′ have larger regions where the modal displacement is near zero (Fig. 1d). Therefore, by attaching the supporting structures to these regions, it is possible to reduce energy dissipation of the ring-shaped mechanical resonator to the substrate.

Next, we investigate a realistic structure in which two supporting rods are added to the azimuthal-symmetry-broken ring-shaped resonator making a wheel-shaped resonator as shown in Figs. 2a and 2b. We need to engineer this structure and analyze the modal coupling to satisfy the Friedrich–Wintgen condition [Eq. (2)] for constructing a mechanical BIC. Figure 2a is a three-dimensional view of the entire device structure where the wheel-shaped resonator is seated on a silicon oxide ($SiO_2$) pedestal on the substrate. Figure 2b shows the top and side views of the entire device, where the wheel-shaped silicon micromechanical resonator, the SiO$_2$ pedestal, and the substrate are marked in blue, black, and gray, respectively. The additional two parameters for the wheel-shaped resonator $d$ and $r_s$ are the supporting rods' width and the center disk radius, respectively. To investigate the influence of the supporting rods on the modal coupling, we simulated the frequencies and mechanical $Q$ factors of mode A′ and mode B′ as a function of semi-major axis $r_x$ for different rod widths $d$, with the results shown in Figs. 2c and 2d. The other geometric parameters are fixed at $r_y = 18.7$ μm, $R = 26.1$ μm, and $r_s = 14.7$ μm. The insets in Fig. 2c show the displacement profiles of the corresponding mechanical modes. It can be found that an anticrossing in the modal frequencies (Fig. 2c) and a drastic variation in the mechanical $Q$ factor of mode A′ (Fig. 2d) occur simultaneously near $r_x = 20.8$ μm, despite a large variation of $d$ from 0.5 to 5 μm. These behaviors indicate that mode A′ becomes a Friedrich–Wintgen BIC[24]. Note that the high-$Q$ Friedrich–Wintgen BIC can be obtained in the wheel-shaped resonator with $d$ as large as several micrometers. One reason is that the hybrid mode A′ has a larger region of near-zero displacement than the original uncoupled wine-glass mode (mode B). Actually, we simulated a series of structures with $d$ varying from 0.5 to 8 μm and collected the $r_x$ value and mechanical $Q$ factor when the BIC is achieved (indicated by the purple circles in Fig. 2d), with the results plotted in Figs. 2e and 2f, respectively. Figure 2f shows that the simulated mechanical $Q$ factor of the BIC can maintain above $10^8$ in such a wide range of $d$ from 0.5 to 8 μm (See Supplementary Fig. S1), demonstrating excellent robustness again variations of the width of the dissipation channel. Compared with conventional mechanical systems which reply on minimized supporting rods[6,7] or surrounding phononic bandgap structures[5] for reducing the clamping loss and achieving high mechanical $Q$ factors, the Friedrich–Wintgen BIC can exist in mechanical resonators with simply designed sturdy supporting structures, which substantially alleviate device fabrication difficulty, facilitate thermalization and heat dissipation, and enable device applications in versatile environments.

**Experimental results**

To measure the mechanical BIC, we fabricated the wheel-shaped micromechanical resonators on a silicon-on-insulator wafer and used optomechanical transduction for detecting their mechanical $Q$ factors. Under the guidance of theoretical analysis and numerical simulation, we varied the parameter $r_x$ for different resonator devices while keeping the following structural parameters fixed: $d = 5$ μm, $r_s = 14.7$ μm, $r_y = 18.7$ μm, and $R = 26.1$ μm. We also fabricated a bus waveguide in close proximity of the resonator for coupling light into and out of the resonator. Figure 3a shows scanning electron microscope images of a fabricated device. Note that the wheel-shaped micromechanical resonators also support optical whispering-gallery modes circulating around their outer periphery, which were employed to detect the thermomechanical vibration of the resonators via optomechanical transduction. The inset of Fig. 3a is a close-up showing the details of the coupling region between the resonator and the bus waveguide. Figure 3b shows the experimental setup for device characterization. Light from a semiconductor tunable laser was sent through a fiber polarization controller (FPC) and a variable optical attenuator (VOA) before reaching the device under test (DUT). The light coupled out of the DUT was sent through an erbium-doped fiber amplifier (EDFA) before being collected by a photodetector (PD), which converted the optical signal into the electrical domain for obtaining the optical transmission or for RF spectral analysis. Figure 3c plots a measured optical transmission spectrum of the resonator, where the dips correspond to the optical whispering-gallery modes in different orders. Figure 3d is a close-up of a dip at ~1558.4 nm, which shows that the loaded and intrinsic optical $Q$ factors are $2.3 \times 10^5$ and $7.7 \times 10^5$, respectively.

To experimentally verify the mechanical BIC, we measured the wheel-shaped micromechanical resonators in a vacuum chamber, which could provide an ambient pressure from $1.0 \times 10^5$ to $6.0 \times 10^{-3}$ Pa for the devices. To obtain the intrinsic mechanical $Q$ factor, we attenuated the laser light by using the VOA such that the dynamic backaction from optomechanical interaction was negligible in the resonator (See Supplementary Note 4). Figure 4a plots the simulated and measured frequency of modes A′ and B′ (modal profiles in Fig. 4a inset) for devices with different $r_x$. The simulated results are extracted directly from the rightmost plot Fig. 2c. The measured results agree well with the simulated results, which confirms the existence of the two modes. Figure 4b plots the mechanical $Q$ factors of modes A′ and B′ as a function of $r_x$ measured at the ambient pressure of $6.0 \times 10^{-3}$ Pa. Mode A′ achieves its maximal mechanical $Q$ factor of 9453 at $r_x = 20.8$ μm, which agrees with the simulated results in Fig. 2d. Therefore, we confirm attainment of the mechanical BIC in our fabricated micromechanical resonators. Figure 4c shows the measured displacement noise power spectral density of modes A′ and B′ at $r_x = 20.8$ μm where the BIC is achieved. The Lorentzian fitting of these spectra provides the mechanical $Q$ factor of 9453 for mode A′ and 882 for mode B′. The measured mechanical $Q$ factor of mode A′ at the BIC point is lower than the simulated value in Fig. 2d. It should be noted that the strategy of engineering a Friedrich–Wintgen BIC can only be used for eliminating the clamping loss. The other loss mechanisms such as air damping loss and material loss cannot be reduced effectively by the structural engineering and modal control[25,26]. Next, we investigated the residual loss in our BIC device ($r_x$ = 20.8 μm) where the clamping loss has been completely eliminated. To this end, we measured its mechanical $Q$ factor under different ambient pressures in the vacuum chamber, with the results shown in Fig. 4d. The mechanical $Q$ factors for devices with other $r_x$ values at different ambient pressures can be found in Supplementary Fig. S2a. Figure 4d shows that the mechanical $Q$ factor decreases as the ambient pressure increases and this effect becomes more pronounced when the ambient pressure is above 1 Pa, which indicates that air damping loss was the main loss mechanism. Since material loss usually depends on temperature and maintains constant at a given temperature, e.g., room temperature in our experiment, we can express the mechanical $Q$ factor as

$$Q^{-1} = Q_0^{-1} + Q_{ext}^{-1} = Q_0^{-1} + C^{-1}P, \tag{5}$$

where $Q_0$, $Q_{ext}$, and $P$ are the intrinsic $Q$ factor, extrinsic $Q$ factor, and the ambient pressure, respectively. $C$ is a proportionality constant defined as $\rho h f \sqrt{\pi^3 RT/8M}$, where $\rho$, $h$, $f$, $R$, $T$, and $M$ are the material mass density, resonator thickness, mechanical frequency, ideal gas constant, temperature, and molar mass of air, respectively[27]. With $\rho$ = 2329 kg m$^{-3}$, $h$ = 220 nm, $f$ = 57 MHz, $R$ = 8.31 J K$^{-1}$ mol$^{-1}$, $T$ = 300 K, and $M$ = 28.97 g mol$^{-1}$, $C$ has the theoretically calculated value of 1.69 × 10$^7$ Pa. Note that the above expression for the mechanical $Q$ factor in Eq. (5) applies only to a relatively low ambient pressure where the free-molecular-flow approximation is valid. Under a high pressure, the air-damping-dominated $Q_{ext}$ follows a $P^{-1/2}$ dependence[27]. Therefore, we fitted the experimental results at the ambient pressure below 10$^4$ Pa based on Eq. (5), obtaining the red curve shown in Fig. 4d with the fitted $C$ being 1.49 × 10$^7$ Pa, which agrees well with the theoretically calculated value.

**Conclusion**

In summary, we experimentally realized a BIC in an individual micromechanical resonator, which provides a new strategy of phonon trapping in micromechanical structures with dissipation channels. By breaking the azimuthal symmetry, we introduced coupling between two dissipative modes of a wheel-shaped micromechanical resonator for making destructive interference between the dissipation channels. As a result, we obtained a Friedrich–Wintgen BIC with zero clamping loss, and achieved a mechanical $Q$ factor of ~10$^4$ in the very high frequency band at room temperature from a wheel-shaped micromechanical resonator with its supporting rods' width as large as 5 μm. To obtain high-$Q$ resonances in individual micromechanical resonators, the conventional wisdom relies on minimizing the size of the supporting structures which renders the fabricated mechanical device fragile. Defying the conventional wisdom, our strategy applies to robust mechanical structures, which not only substantially reduces device fabrication difficulty but also enables device operation in versatile environments for broader application areas. Our experimental results open a new way of obtaining high-$Q$ micro- and nanomechanical resonators, which will inspire plenty of applications in interdisciplinary research areas such as electromechanics, optomechanics, and quantum physics.

**Online Content** Any methods, additional references, Nature Research reporting summaries, source data, statements of data availability and associated accession codes are available in the online version of the paper; references unique to these sections appear only in the online paper.

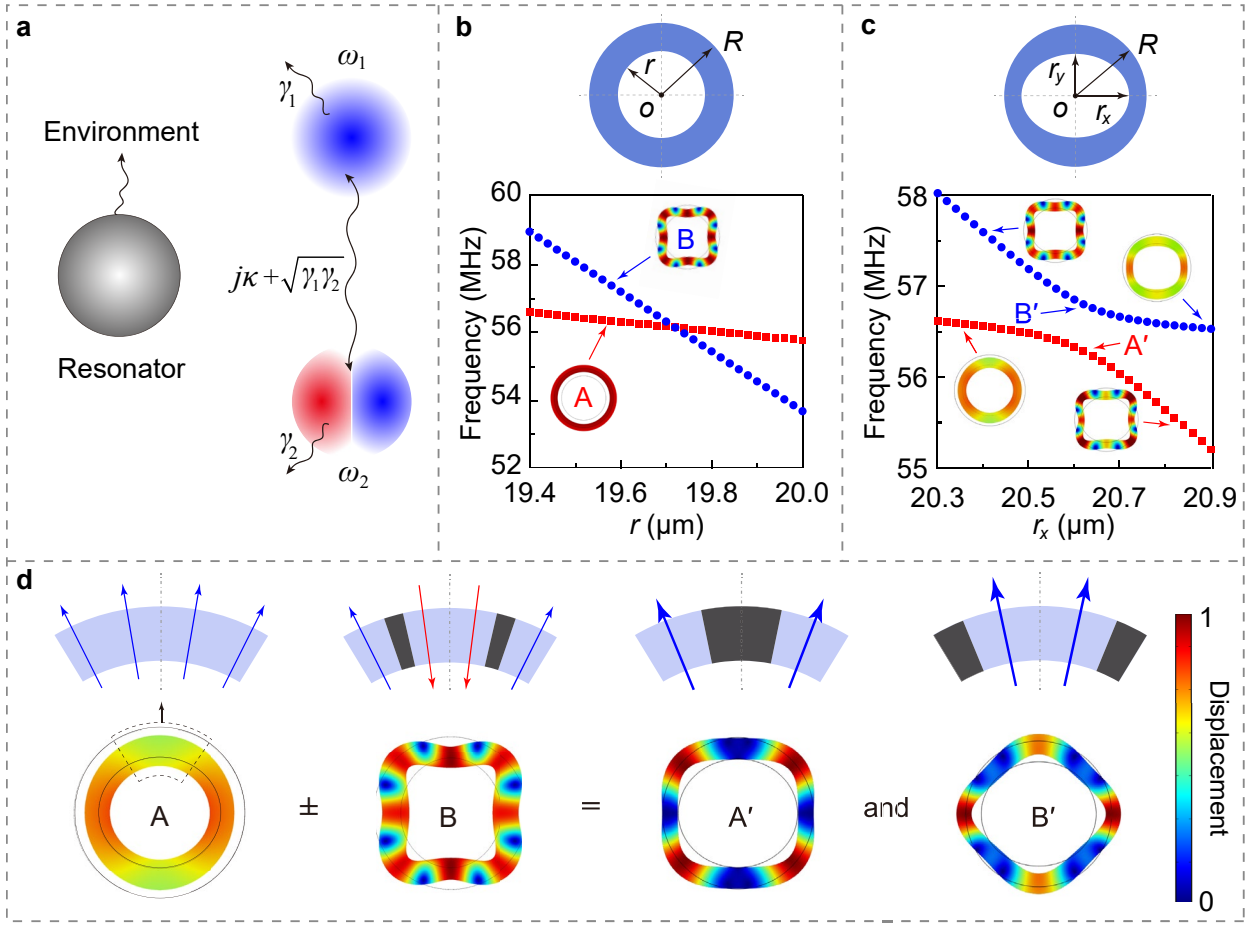

**Fig. 1 | Construction of BIC in a micromechanical resonator based on modal coupling. a**, Schematic of dispersive and dissipative coupling between two eigenmodes of a resonating system. **b**, Simulated frequencies of mode A (red rectangles) and mode B (blue dots) of a ring-shaped silicon micromechanical resonator with azimuthal symmetry as a function of the inner radius $r$. The resonator has a thickness $h$ = 220 nm and outer radius $R$ = 26.1 μm. Modes A and B are the fundamental radial-contour mode and the 4th-order wine-glass mode, respectively. **c**, Simulated frequencies of mode A′ (red rectangles) and mode B′ (blue dots) of a ring-shaped resonator with broken azimuthal symmetry as a function of semi-major axis of the inner boundary $r_x$. The resonator has a thickness $h$ = 220 nm, outer radius $R$ = 26.1 μm, and semi-minor axis of the inner boundary $r_y$ = 18.7 μm. **d**, Generation of the hybrid modes A′ and B′ from coupling of the original modes A and B at the anticrossing point.

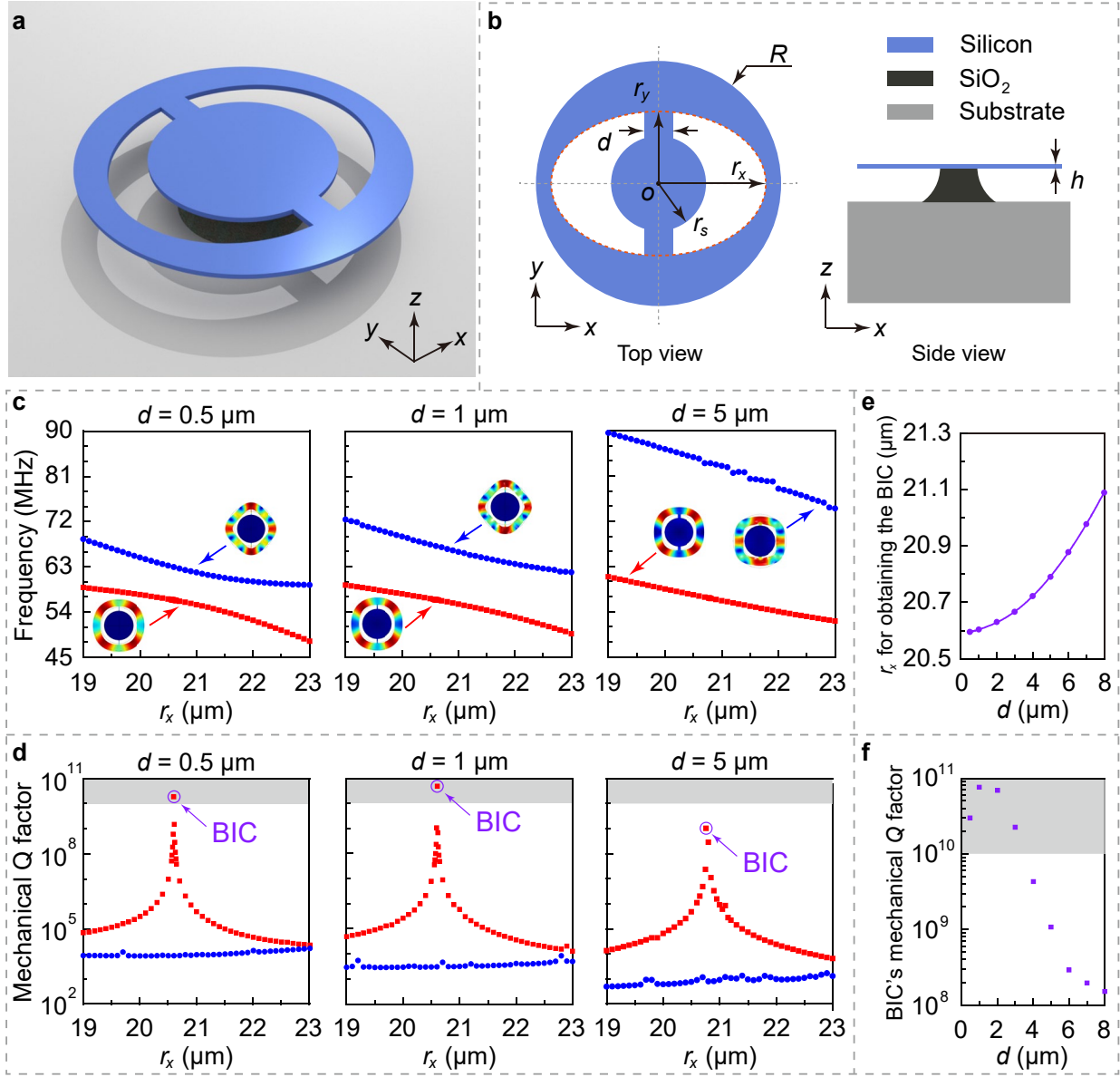

**Fig. 2 | Design and numerical simulation of BIC in a micromechanical resonator. a**, Illustration of the wheel-shaped micromechanical resonator that supports BIC. **b**, Top and side views of the micromechanical resonator, with dimension labels. **c**, Simulated frequencies of modes A′ and B′ of the micromechanical resonator as a function of $r_x$ for $d$ = 0.5, 1, and 5 μm. The insets show the corresponding modal displacement profiles. **d**, Simulated mechanical $Q$ factors of modes A′ and B′ as a function of $r_x$ for $d$ = 0.5, 1, and 5 μm. The BIC point where the highest mechanical $Q$ factor is achieved is marked in each plot. **e**, The value of $r_x$ for obtaining the BIC as a function of $d$. **f**, Simulated mechanical $Q$ factor of the BIC as a function of $d$. In **c**–**f**, the fixed geometric parameters of the micromechanical resonator are $r_y$ = 18.7 μm, $R$ = 26.1 μm, $r_s$ = 14.7 μm, and $h$ = 220 nm. In **d** and **f**, the gray areas indicate the regimes reaching the limit of numerical simulation, where the simulated results do not converge.

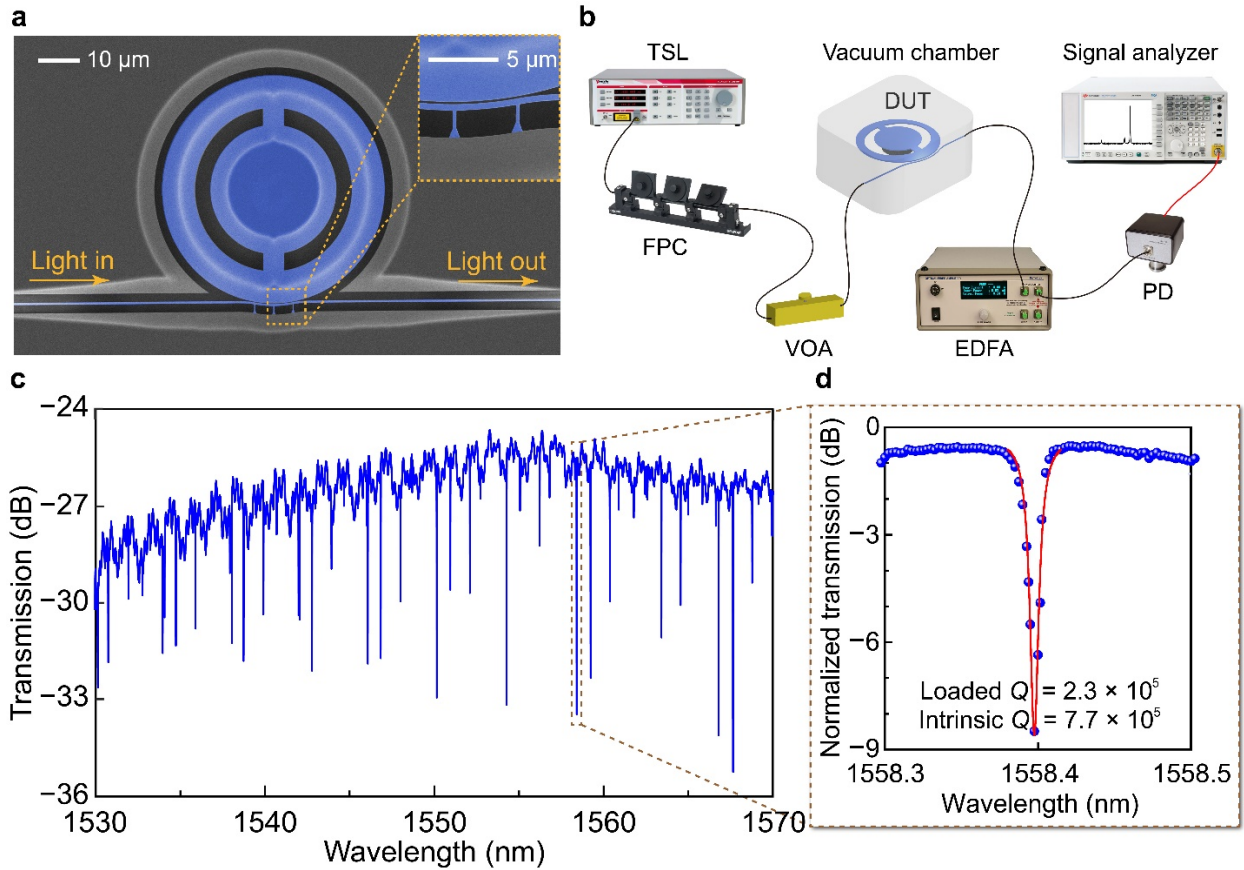

**Fig. 3 | Device fabrication and experimental characterization**. **a**, Scanning electron microscope image of a fabricated silicon micromechanical resonator. The nearby bus waveguide is used for coupling light into the resonator for optical measurement of its mechanical modes. The inset is a close-up showing the details of the coupling region between the resonator and the bus waveguide. **b**, Experimental setup. TSL, tunable semiconductor laser; FPC, fiber polarization controller; VOA, variable optical attenuator; DUT, device under test; EDFA, erbium-doped fiber amplifier; PD, photodetector. **c**, Measured optical transmission spectrum of the device in **a**. **d**, Zoomed-in optical transmission spectrum showing an optical resonance with Lorentzian-fitted optical $Q$ factors.

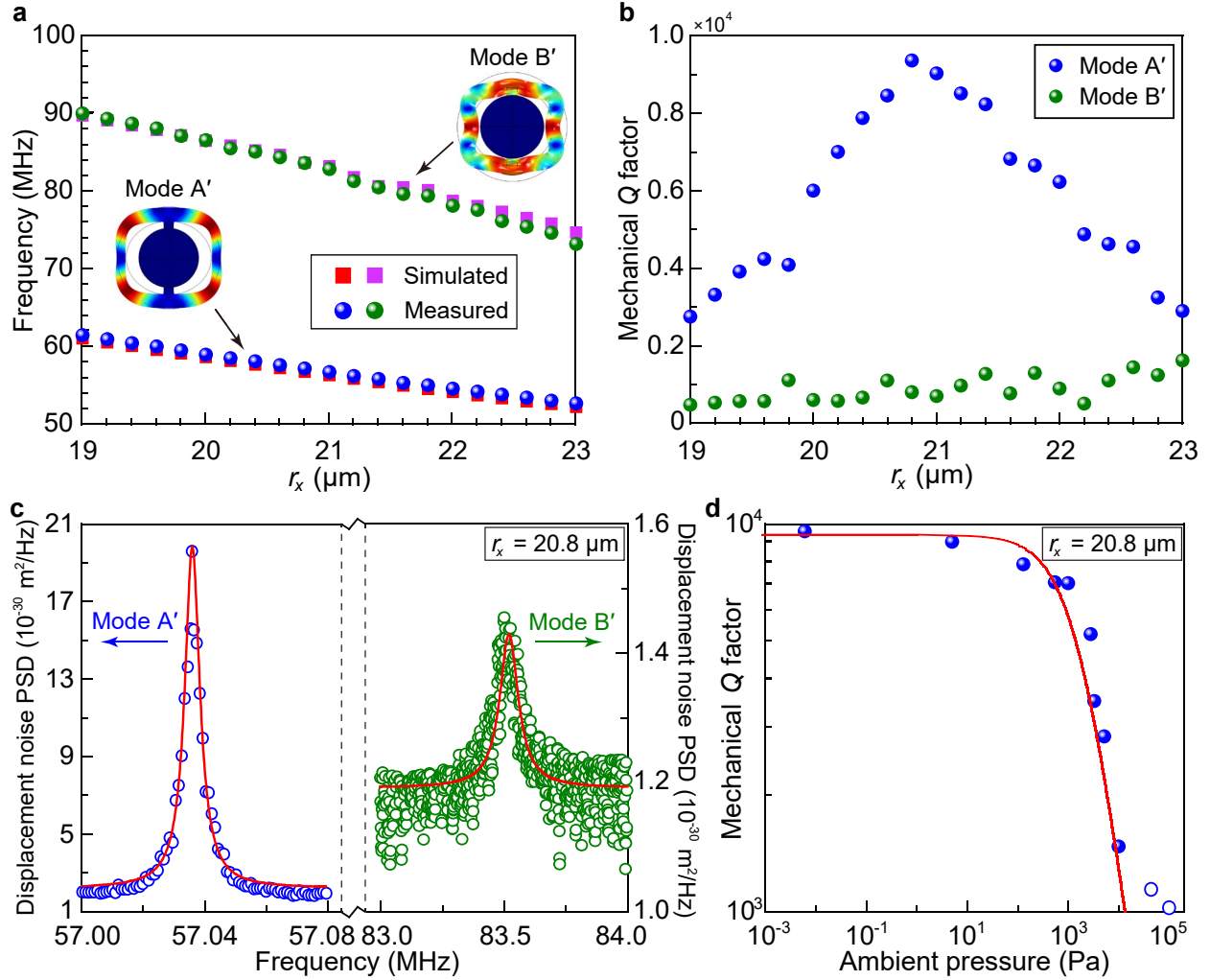

**Fig. 4 | Experimental demonstration of BIC in a micromechanical resonator. a**, Simulated and measured frequencies of modes A′ and B′ as a function of $r_x$. **b**, Measured mechanical $Q$ factors of modes A′ and B′ as a function of $r_x$ under the ambient pressure of $6.0 \times 10^{-3}$ Pa. **c**, Measured displacement noise power spectral density (PSD) of modes A′ and B′ from the device with $r_x$ = 20.8 μm under the ambient pressure of $6.0 \times 10^{-3}$ Pa. The blue (green) open circles represent the measured data points for mode A′ (B′), and the red lines are the corresponding Lorentzian fits. **d**, Measured mechanical $Q$ factor of mode A′ from the device with $r_x$ = 20.8 μm as a function of the ambient pressure. The data collected below (above) $10^4$ Pa are marked in blue dots (open circles). The red curve plots a theoretical fit for the blue data points (below $10^4$ Pa) based on Eq. (5).